\documentclass[12pt]{iopart}
\usepackage{iopams}
\expandafter\let\csname equation*\endcsname\relax
\expandafter\let\csname endequation*\endcsname\relax
\usepackage{amsmath}
 
\begin{document}
\title{Note on the transformation from the Lorenz gauge to the Coulomb gauge}
\author{V Hnizdo$^1$ and G Vaman$^2$} 
\address{$^1$ 2044 Georgian Lane, Morgantown, WV 26508, USA}
\address{$^2$ Aleea Callatis 1, Bucharest, Romania}
\eads{\mailto{hnizdo2044@gmail.com} and \mailto{getavaman@gmail.com}}

\begin{abstract}
There is a simple formula for the transformation from the Lorenz gauge to the Coulomb gauge, valid under a condition that is satisfied by some charge densities employed in the literature. An equation for the gauge function of this transformation that is alternative but equivalent to a formula of Jackson is derived also. 
\end{abstract}


\noindent
Jackson \cite{Jack} has derived formulae for the gauge function $\chi_{\rm C}({\bi r},t)$ that transforms  Lorenz-gauge potentials $\Phi_{\rm L}({\bi r},t)$ and ${\bi A}_{\rm L}({\bi r},t)$ to the equivalent Coulomb-gauge potentials
$\Phi_{\rm C}({\bi r},t)$ and ${\bi A}_{\rm C}({\bi r},t)$ according to (Gaussian units are used):
\begin{align}
\partial\chi_{\rm C}({\bi r},t)/c\partial t= \Phi_{\rm L}({\bi r},t)-\Phi_{\rm C}({\bi r},t), \quad
\bnabla\chi_{\rm C}({\bi r},t)= {\bi A}_{\rm C}({\bi r},t)-{\bi A}_{\rm L}({\bi r},t).
\label{def}
\end{align}
One of Jackson's formulae was used in \cite{VH1}
to calculate the gauge function of the transformation of the Lorenz-gauge  potentials of a uniformly moving point charge to the Coulomb gauge. 
This formula reads
\begin{align}
\chi_{\rm C}({\bi r},t)=-c\int \rmd^3 r'\frac{1}{R}\int_0^{R/c}\rmd\tau\,\rho({\bi r}',t-\tau) +\chi_0,
\label{chiCJack}
\end{align}
where $R=|{\bi r}-{\bi r}'|$, $\rho({\bi r},t)$ is the charge density generating  the scalar potentials 
$\Phi_{\rm L}$ and $\Phi_{\rm C}$, and 
$\chi_0$ is an integration term that Jackson shows to be, remarkably, at most
a constant independent of $\bi r$ and $t$.
 
In this Note, we show that there is a simpler-to-use formula when the `instantaneous,' and so usually easily calculable,  scalar potential $\Phi_{\rm C}$ is known, and the requisite  charge density  satisfies the condition
\begin{align}
\lim_{t\rightarrow -\infty}\int \rmd^3 r'\frac{1}{R}\int_{t-R/c}^t \rmd t'\rho({\bi r}',t') = {\rm const},
\label{cond}
\end{align}
where const is a quantity independent of $\bi r$ and $t$.
This is seen as follows. The integration of
the first equality in equation (\ref{def}) with respect to time from  $-\infty$ to a time $t$  yields
\begin{align} 
\chi_{\rm C}({\bi r},t)=c\int_{-\infty}^t \rmd t'\,[\Phi_{\rm L}({\bi r},t')-\Phi_{\rm C}({\bi r},t')]
+\chi_{\rm C}({\bi r},-\infty),
\label{chiC1}
\end{align}
where $\chi_{\rm C}({\bi r},-\infty)\equiv\lim_{t\rightarrow -\infty}\chi_{\rm C}({\bi r},t)$,
which is assumed to exist.
On a close reading of Jackson's paper \cite{Jack}, it can be seen that the integration term 
$\chi_{\rm C}({\bi r},-\infty)$ is
expressible in terms of the charge density $\rho({\bi r},t)$ as
\begin{align} 
\chi_{\rm C}({\bi r},-\infty)=\lim_{t_0\rightarrow -\infty}\left[-c\int \rmd^3 r'\frac{1}{R}\int_{t_0-R/c}^{t_0}\rmd t\,\rho({\bi r}',t)\right]+\chi_0,
\label{intterm}
\end{align} 
where  $t_0$ is the lower limit of Jackson's time integration and $\chi_0$ is the same constant as in equation (\ref{chiCJack}).\footnote
{Jackson integrates his Eq.\,(3.3), which expresses $\partial\chi_{\rm C}/c\partial t$ in terms of integral representations of $\Phi_{\rm L}$ and $\Phi_{\rm C}$,
from $t'=t_0$ to $t$.  This yields 
$\chi_{\rm C}({\bi r},t)-\chi_{\rm C}({\bi r},t_0) =c\int \rmd^3 r'\frac{1}{R}[\int_{t_0-R/c}^{t-R/c} \rmd t'\rho({\bi r}',t')-\int_{t_0}^t \rmd t'\rho({\bi r}',t')]$, which can be written as his Eq.\,(3.4),
$\chi_{\rm C}({\bi r},t) =c\int \rmd^3 r'\frac{1}{R}[\int_{t_0}^{t-R/c}\rmd t'\rho({\bi r}',t')-\int_{t_0}^{t}\rmd t'\rho({\bi r}',t')]+ \chi_0$, when $\chi_{\rm C}({\bi r},t_0)=
-c\int \rmd^3 r'\frac{1}{R}\int_{t_0-R/c}^{t_0}\rmd t'\rho({\bi r}',t')+\chi_0$.}
Under the condition ({\ref{cond}), 
the integration term (\ref{intterm}) reduces to a  constant, 
simplifying equation (\ref{chiC1}) to read 
\begin{align}
\chi_{\rm C}({\bi r},t)= F({\bi r},t)-\lim_{t\rightarrow -\infty}F({\bi r},t)+{\rm const},
\label{chiC2}
\end{align}
where 
\begin{align}
F({\bi r},t) =c\int \rmd t\,[\Phi_{\rm L}({\bi r},t)-\Phi_{\rm C}({\bi r},t)],
\label{F}
\end{align}
which is an indefinite integral with respect to time.  The constant  in  (\ref{chiC2})  can be omitted since a gauge function is defined only to within an additive constant. Equation (\ref{chiC2}) with equation (\ref{F}) furnish the desired formula.
 
Condition (\ref{cond}) is satisfied in the case of a uniformly moving point charge, say case 1, and also in its more involved variant, in which the charge is set, suddenly, into its uniform motion from rest, case 2.
The charge densities of cases 1 and 2 are, respectively,
\begin{align}
\rho_1({\bi r},t)&=q\,\delta(x-vt)\delta(y)\delta(z),
\label{rho1} \\
\rho_2({\bi r},t)&=q\,\delta(x-vt)\delta(y)\delta(z)\Theta(t)+q\,\delta(x)\delta(y)\delta(z)\Theta(-t),
\label{rho2}
\end{align}
where $v\hat{\bi x}$ is the charge's velocity and it is assumed that the charge passes or starts to move from the origin ${\bi r}=0$ at a time $t=0$; $\Theta(\cdot)$ is the Heaviside step function. 
These densities satisfy the condition (\ref{cond}) since
\begin{align}
\lim_{t_0\rightarrow -\infty}\left[-c\int \rmd^3 r'\frac{1}{R}\int_{t_0-R/c}^{t_0} \rmd t\,\rho_1({\bi r}',t)\right] &=
\frac{q}{|\beta|}\, \ln(1-|\beta|),\quad \beta=\frac{v}{c},\label{condrho1}\\
\lim_{t_0\rightarrow -\infty}\left[-c\int \rmd^3 r'\frac{1}{R}\int_{t_0-R/c}^{t_0} \rmd t\,\rho_2({\bi r}',t)\right] &=-q.
\label{condrho2}
\end{align}
The calculations of the results (\ref{condrho1}) and (\ref{condrho2}) are given in Appendix.

Using formula (\ref{chiC2}) for $\chi_{\rm C}$  in case 1 yields
\begin{align}
\chi_{\rm C\,1}({\bi r},t)= \frac{q}{\beta}\,\left[{\rm arsinh}\frac{x-vt}{\sqrt{y^2+z^2}}-{\rm arsinh}
\frac{\gamma(x-vt)}{\sqrt{y^2+z^2}}\right],
\label{chiC3}
\end{align}
where $\gamma=(1-\beta^2)^{-1/2}$, which simplifies the expression for $\chi_{\rm C}$ obtained for this case in  \cite{VH1}.\footnote{See Eq.\,(13) of \cite{VH1}, in which ${\rm arsinh}[(x-x_0)/\sqrt{y^2+z^2}] 
={\rm arsinh}[\gamma(x-vt)/\sqrt{y^2+z^2}]+{\rm arsinh}(\beta\gamma)$, which can be shown to hold with  $x_0$ as given there by  Eq.\,(12) [note that the quantity $x_0$ then satisfies
$\sqrt{(x-x_0)^2+y^2+z^2}=(1/\beta)(vt-x_0)$].}
The expression on the RHS of equation (\ref{chiC3}) is the indefinite integral  (\ref{F}), where
\begin{align}
\Phi_{\rm L\, 1}({\bi r},t)&=\frac{q}{\sqrt{(x-vt)^2+(y^2+z^2)/\gamma^2}},
\label{PhiL1} \\
\Phi_{\rm C\, 1}({\bi r},t)&=\frac{q}{\sqrt{(x-vt)^2+y^2+z^2}}
\end{align}
are the requisite scalar potentials \cite{VH1}; its limit 
$t\rightarrow -\infty$ is a constant equal to
$-(q/|\beta|){\rm arsinh}(\beta^2\gamma/2)$ 
and as such is omitted in (\ref{chiC3}).

For case 2,  formula  (\ref{chiC2}) yields  a gauge function
\begin{align}
\chi_{\rm C\,2}({\bi r},t)&=\frac{q}{\beta}\,\left[{\rm arsinh}\frac{\gamma(x-\beta r)}{\sqrt{y^2+z^2}}-{\rm arsinh}\frac{\gamma(x-vt)}{\sqrt{y^2+z^2}}\right]\Theta(ct-r)\nonumber\\
&\quad+\frac{q}{\beta}\,\left[{\rm arsinh}\frac{x-vt}{\sqrt{y^2+z^2}}-{\rm arsinh}\frac{x}{\sqrt{y^2+z^2}}\right]
\Theta(t)\nonumber\\
&\quad+\frac{q}{r}[(r-ct)\Theta(ct-r)+ct\Theta(t)].
\label{chiC4} 
\end{align} 
The expression on the RHS of equation (\ref{chiC4}) is again the indefinite integral  (\ref{F}),
where the requisite scalar potentials are now
\begin{align}
\Phi_{\rm L\, 2}({\bi r},t)&=\Phi_{\rm L\, 1}({\bi r},t)\,\Theta(ct-r)+(q/r)\,\Theta(r-ct), \label{PhiL2}\\
\Phi_{\rm C\, 2}({\bi r},t)&=\Phi_{\rm C\, 1}({\bi r},t)\,\Theta(t)+(q/r)\,\Theta(-t) \label{PhiC2}.
\end{align}
At any $ t<0$, the gauge function (\ref{chiC4}) vanishes.
Recently, case 2 has been  considered in \cite{HV}, where the Lorenz-gauge scalar potential (\ref{PhiL2})
is deduced and the same result for the gauge function $\chi_{\rm C}$ as that of equation (\ref{chiC4}) is  obtained and validated.\footnote{Regrettably, in  the sentence just below  Eq.\,(40) of \cite{HV},  
the integration term $\chi_{\rm C}({\bi r},t_0)$ is confused with Jackson's term $\chi_0$.  
Irrespective of that,
the calculation of the gauge function $\chi_{\rm C}$ in \cite{HV} is in accordance with formula (\ref{chiC2}) of this Note.}

In closing, we note that an equation for 
$\chi_{\rm C}({\bi r},t)$ that is alternative but equivalent to Eq.\,(\ref{chiCJack}) can be obtained as a solution to the Poisson equation the gauge function 
satisfies, using a procedure similar to that employed by Jackson in his calculation of an auxiliary function $\Psi$ in [1, Sec.\,IV]. The procedure is outlined by the following equations:
\begin{align}  
\nabla^2 \chi_{\rm C}({\bi r},t)&= \frac{\partial\Phi_{\rm L}({\bi r},t)}{c\partial t},\\
\Phi_{\rm L}({\bi r},t)&=\int \frac{\rmd^3 r'}{|{\bi r}-{\bi r}'|}\,
\rho({\bi r}', t-|{\bi r}-{\bi r}'|/c),\\
\chi_{\rm C}({\bi r},t)&= -\frac{1}{4\pi c}\int \frac{\rmd^3 r''}{|{\bi r}-{\bi r}''|}\frac{\partial}{\partial t}\,\int\frac{\rmd^3 r'}{|{\bi r}''-{\bi r}'|}\,\rho({\bi r}', t-|{\bi r}''-{\bi r}'|/c)\nonumber\\
&=-\frac{1}{4\pi c}\int\rmd^3 r'\int\rmd^3 r''\frac{1}{|{\bi r}-{\bi r}''||{\bi r}''-{\bi r}'|}\frac{\partial\rho({\bi r}', t-|{\bi r}''-{\bi r}'|/c)}{\partial t},\\
&({\bi r}',{\bi r}'')\rightarrow({\bi r}',{\bi R}'={\bi r}''{-}{\bi r}'),\,\rmd^3r''=\rmd^3R',\, {\bi R}={\bi r}-{\bi r}',\\
\chi_{\rm C}({\bi r},t)&=-\frac{1}{4\pi c}\int\rmd^3 r'\int\rmd^3 R'\frac{1}{|{\bi R}-{\bi R}'|R'}
\frac{\partial\rho({\bi r}', t-R'/c)}{\partial t},\label{chiCHV1}\\
\frac{1}{|{\bi R}-{\bi R}'|}&=\sum_{l=0}^{\infty}\frac{R_<^l}{R_>^{l+1}}\,P_l(\hat{\bi R}\bdot\hat{\bi R}'),\,{\rm only}\;l=0\;{\rm term\;contributes\;to\;(\ref{chiCHV1})},
\\
\frac{\partial\rho({\bi r}', t{-}R'/c)}{\partial t}&=-c\, \frac{\partial\rho({\bi r}', t{-}R'/c)}{\partial R'},\\
\chi_{\rm C}({\bi r},t)&=\int\rmd^3 r'\left[\frac{1}{R}\int_0^R\rmd R'R'\frac{\partial\rho({\bi r}', 
t{-}R'/c)}{\partial R'}+\int_R^{\infty}\rmd R'\frac{\partial\rho({\bi r}', t{-}R'/c)}{\partial R'}\right]
\nonumber\\
&=-\int \rmd^3 r'\frac{1}{R}\int_0^R \rmd R'\rho({\bi r}', t{-}R'/c)+\lim_{R'\rightarrow\infty}\int \rmd^3 r'\rho({\bi r}',t{-}R'/c) \nonumber \\
&=-c\int \rmd^3 r'\frac{1}{R}\int_0^{R/c} \rmd \tau\rho({\bi r}', t{-}\tau)+q.
\end{align}

Observe  that, unlike in Eq.\,(\ref{chiCJack}), in the resulting equation for $\chi_{\rm C}({\bi r},t)$,
\begin{align}
\chi_{\rm C}({\bi r},t)=-c\int \rmd^3 r'\frac{1}{R}\int_0^{R/c} \rmd \tau\rho({\bi r}', t{-}\tau)+q,
\label{chiCHV2}
\end{align}
the 2nd term on the RHS  is  explicitly a non-arbitrary  constant, equal to the total charge $q$ of the charge density $\rho({\bi r},t)$.

\section*{Appendix}
\setcounter{equation}{0}
\renewcommand{\theequation}{A\arabic{equation}}
The integral in (\ref{condrho1}) can be reduced to a one-dimensional integral,
\begin{align}
-c&\int \rmd^3 r'\frac{1}{R}\int_{t_0-R/c}^{t_0}\rmd t\,q\,\delta(x'-vt)\delta(y')\delta(z') \nonumber \\
& =
-\frac{q }{\beta}\int_{-\infty}^{\infty}\frac{ \rmd x'}{\sqrt{(x-x')^2+y^2+z^2}}\int_{t_0-\frac{1}{c}\sqrt{(x-x')^2+y^2+z^2}}^{t_0}\rmd t\,\delta(t-x'/v)\nonumber \\
&=-\frac{q }{\beta}\int_{-\infty}^{vt_0}\rmd x'\,\frac{\Theta(x'-vt_0+\beta\sqrt{(x-x')^2+y^2+z^2})}{\sqrt{(x-x')^2+y^2+z^2}}.
\label{intrho1}
\end{align}
It is  assumed here that $v>0$. 

The limit $t_0\rightarrow -\infty$ of the integral (\ref{intrho1}) is evaluated as follows.
The condition set in the integrand in (\ref{intrho1}) by the Heaviside step function,
\begin{align}
\beta\sqrt{(x-x')^2+s^2}>vt_0-x'>0,\quad s=\sqrt{y^2+z^2},
\end{align}
leads to
\begin{align}
(1/\gamma^2)x'^2+2(\beta^2x-vt_0)x'+v^2t_0^2-\beta^2r^2<0,
\end{align}
which for $t_0\rightarrow -\infty$ simplifies to
\begin{align}
(1/\gamma^2)x'^2-2vt_0x'+v^2t_0^2<0.
\label{binomial}
\end{align}
The roots of the LHS of  (\ref{binomial}) are
\begin{align}
x_1'=\gamma^2vt_0(1-\beta),\quad x_2'=\gamma^2vt_0(1+\beta),
\end{align}
for which a negative $vt_0$ satisfies $x_2'<vt_0<x_1'$.
The integral in  (\ref{intrho1}) then evaluates for  $t_0\rightarrow -\infty$ as
\begin{align}
\int_{x_2'}^{vt_0}\frac{\rmd x'} {\sqrt{(x-x')^2+s^2}} &={\rm arsinh}\frac{x'-x}{s}\,\Big |_{x_2'}^{vt_0}
\nonumber \\
&={\rm arsinh}\left[\frac{vt_0-x}{s^2}\sqrt{s^2+(x_2'-x)^2}
-\frac{x_2'-x}{s^2}\sqrt{s^2+(vt_0-x)^2}\right]\nonumber \\
&={\rm arsinh}\left[\frac{(vt_0-x)|x_2'-x|}{s^2}\left(1+\frac{s^2}{2(x_2'-x)^2}\right)\right.\nonumber\\
&\left.\quad\quad\quad\quad-\frac{(x_2'-x)|vt_0-x|}{s^2}\left(1+\frac{s^2}{2(vt_0-x)^2}\right)
+{\rm O}(s^2/v^2t_0^2)\right]
\nonumber\\
&=-{\rm arsinh}\left[\frac{vt_0-x}{2(x_2'-x)}-\frac{x_2'-x}{2(vt_0-x)}
+{\rm O}(s^2/v^2t_0^2)\right].
\label{resultrho1}
\end{align}
Here, 
in the last line, $|x_2'-x|$ and $|vt_0-x|$ are set to equal $-(x_2'-x)$ and $-(vt_0-x)$, respectively, in view of $t_0\rightarrow -\infty$.  

Using  equation (\ref{resultrho1}), we obtain
\begin{align}
&\lim_{t_0\rightarrow -\infty}\left[-c\int \rmd^3  r'\frac{1}{R} \int_{t_0-R/c}^{t_0}\rmd t \,q\,\delta(x'{-v}t)\delta(y')\delta(z')\right]\nonumber \\ 
&\quad\quad\quad=\lim_{t_0\rightarrow -\infty}\frac{q}{\beta}\,{\rm arsinh}\left[\frac{vt_0-x}{2(x_2'-x)}-\frac{x_2'-x}{2(vt_0-x)}\right]\nonumber \\
&\quad\quad\quad=\frac{q}{\beta}\,{\rm arsinh}\left[\frac{1}{2}\left(1-\beta-\frac{1}{1-\beta}\right)\right]\nonumber \\
&\quad\quad\quad=\frac{q}{\beta}\,\ln(1-\beta).
\label{positivev}
\end{align}
A similar calculation for $v<0$ leads to the generalization of the result (\ref{positivev}) for a $\beta$ of either sign,  
\begin{align}
\lim_{t_0\rightarrow -\infty}&\left[-c\int \rmd^3 r'\frac{1}{R}\int_{t_0-R/c}^{t_0}\rmd t\,q\,\delta(x'-vt)\delta(y')\delta(z')\right]=\frac{q}{|\beta|}\,\ln(1-|\beta|).
\end{align}

The calculation of the result (\ref{condrho2}) turns out to be much simpler. We have
\begin{align}
&-c\int \rmd^3 r'\frac{1}{R}\int_{t_0-R/c}^{t_0}\rmd t\,q[\delta(x'-vt)\delta(y')\delta(z')\Theta(t)
+\delta({\bi r}')\Theta(-t)] \nonumber \\
&\quad\quad= 
-\frac{q}{|\beta|}\int_{-\infty}^{\infty}\frac{\rmd x'}{\sqrt{(x-x')^2+y^2+z^2}}\int_{t_0-\frac{1}{c}\sqrt{(x-x')^2+y^2+z^2}}^{t_0}\rmd t\,\delta(t-x'/v)\Theta(t)\nonumber \\
&\quad\quad\quad-\frac{qc}{r}\int_{t_0-r/c}^{t_0}\rmd t\,\Theta(-t).
\label{intrho2}
\end{align}
When  $t_0 <0$,  the integral in question  equals just the expression in the last line of equation (\ref{intrho2}). So, 
\begin{align}
&\lim_{t_0\rightarrow -\infty}\left[-c\int \rmd^3 r'\frac{1}{R}\int_{t_0-R/c}^{t_0}\rmd t\,q[\delta(x'-vt)\delta(y')\delta(z')\Theta(t)+\delta({\bi r}')\Theta(-t)]\right]
\nonumber \\
&\quad\quad\quad=
\lim_{t_0\rightarrow -\infty}\left[-\frac{qc}{r}\,\int_{t_0-r/c}^{t_0}\rmd t\,\Theta(-t)\right]
=\lim_{t_0\rightarrow -\infty}\left[-\frac{qc}{r}\,(t_0-t_0+r/c)\right]
\nonumber\\
&\quad\quad\quad=-q.
\end{align}

\end{document}